# HIDDEN GRAVITY IN
# OPEN-STRING FIELD THEORY

W. Siegel[1]

*Institute for Theoretical Physics*
*State University of New York, Stony Brook, NY 11794-3840*

**ABSTRACT**

We clarify the nature of the graviton as a bound state in open-string field theory: The flat metric in the action appears as the vacuum value of an *open* string field. The bound state appears as a composite field in the *free* field theory.

---

[1]　Internet address: siegel@insti.physics.sunysb.edu.

# 1. INTRODUCTION

Fundamental strings (as opposed to hadronic strings) have been proposed as the solution to two problems — unified theory and quantum gravity. The use of strings as a solution to the former problem hinges on its use as a solution to the latter: Since the compactifications of string theories to four dimensions are so numerous (and the compactification itself does not seem to be predictable), it is not clear that the predictability of string theory for low energy physics is much greater than that of renormalizable, or low-energy phenomenological, four-dimensional quantum field theory (of particles). The greater predictability of string theory is expected from the (hopefully-not-too-much) higher-energy corrections arising from the direct or indirect effects of gravity (including the effects of supersymmetry, whose breaking is best explained through the use of supergravity and the superHiggs effect). In other words, the unification powers of string theory, although originally thought to be great because of the few models available in ten dimensions before compactification, have been reduced to a solution to the unification of gravity with (super) grand unified theories.

Therefore string theory is basically a solution only to the problem of quantum gravity, but it is the only known solution. Ordinarily that might be sufficient (and perhaps even desirable), but until the experimental situation improves, it would be useful to have an alternative theory of quantum gravity for purposes of comparison. The only proposed mechanism free of ghosts (and the resultant effectively nonrenormalizable ambiguities) is the appearance of the graviton as a bound state in a renormalizable field theory. One advantage this might have over string theory is the correct prediction of the dimensionality of space time: (super)string theory has critical dimension 10, while renormalizable field theory (with a finite number of fields and bounded potential) has critical (maximum) dimension 4. However, the only known theory in which the graviton has been demonstrated to appear as a bound-state pole in the S-matrix is open-string theory [1]. Therefore the phenomenon of bound-state gravity in open-string theory warrants a closer study, with an eye toward isolating those aspects that are essential to this phenomenon but might not require string theory.

In this paper we point out two curious features of this mechanism in open-string theory that may be crucial to the understanding (if not the formulation) of bound-state gravity in a more general setting: (1) In the following section we discuss the nature of the vacuum value of the open-string field. Originally this was introduced as an analogy to gravity where, unlike most other field theories, there is no kinetic (quadratic) term in the action until after expanding about the flat metric. However, in classical open-string field theory (as opposed to classical closed-string field theory) there is no gravity. We discuss the origin of this phenomenon from the vacuum values of massive spin-two fields, the relation of their Stueckelberg fields to the Skyrme model, and the relation of these massive fields to the massless graviton in the quantum theory. (2) In section 3 we study the generation of the graviton at the quantum level. In open-string field theory the graviton appears in a one-loop diagram, rather than through an infinite sum of (one-particle-irreducible) graphs as in most other known theories with bound states (such as QED and QCD). We point out that this implies that the bound-state graviton actually appears in free open-string field theory. We discuss the analogy to the theory of a free, two-dimensional, massless spinor, which has a massless scalar as a bound state. The bound-state graviton in open-string field theory is thus a higher-dimensional analog of bosonization. Besides this analogy between free theories, we also describe the analogy between interacting open-string field theory and the Schwinger model, which shows how these bound states manifest themselves as new poles in the fundamental fields.

It is useful to analyze this phenomenon from the point of view of an effective theory where a new, redundant field is introduced for the bound state, and we therefore discuss the gauge-invariant field theory of coupled open and closed strings in section 4. In the final section we give a more general analysis of coupled systems at the free level, which allows us to discuss some examples of the mixing that occurs between states of the open string and the closed string: the Higgs mechanism for the massless fields of the U(N) string, and the coupling of the graviton of the closed string to the lowest-mass spin-2 field of the open string.

## 2. THE METRIC TENSOR AS AN OPEN-STRING FIELD

The action for open-string field theory [2] has been rewritten as a single term cubic in the fields [3], which is apparently independent of the flat-space metric. The flat-space metric appears through the vacuum expectation value responsible for generating the kinetic term from the current $\mathcal{Q}$ of the BRST operator $Q$:

$$\tilde{\Psi}[X^m(\sigma), C^0(\sigma), C^1(\sigma)] = \langle \tilde{\Psi} \rangle + \Psi$$

$$\Rightarrow \quad S = \int \tfrac{1}{3} \tilde{\Psi}^\dagger \tilde{\Psi} * \tilde{\Psi} = \int (\tfrac{1}{2} \Psi^\dagger Q \Psi + \tfrac{1}{3} \Psi^\dagger \Psi * \Psi)$$

$$\langle \tilde{\Psi} \rangle = Q_L I, \quad Q_L = \int_0^{\pi/2} d\sigma\, \mathcal{Q}, \quad Q = \int_0^{\pi} d\sigma\, \mathcal{Q}$$

$$\mathcal{Q} = C^1(X'^m P_m + C'^0 B_0 + C'^1 B_1) + C^0(\tfrac{1}{2}\eta_{mn}X'^m X'^n + \tfrac{1}{2}\eta^{mn}P_m P_n + C'^0 B_1 + C'^1 B_0)$$

where $I$ is the identity element for the $*$ product (the Sp(2)-invariant vacuum, up to a BRST transformation) and $(P_m, B_0, B_1)$ are the momenta conjugate to $(X^m, C^0, C^1)$. Strominger [4] has explained the appearance of the flat-space metric in terms of the coupling of the open string to a background closed string, in a desire to enlarge the space described by the open-string field to include independent open- and closed-string component fields (or even to describe only closed-string fields). Note that neither the flat-space nor any other metric appears in the definition of the Hilbert-space inner product (functional integration measure) $\langle \Psi_1 | \Psi_2 \rangle = \int \Psi_1^\dagger \Psi_2$: The string field theory formulation is one based on first quantization, and thus uses the Zinn-Justin-Batalin-Vilkovisky (ZJBV) formalism [5]. This means that nonvanishing inner products exist only between fields and their ZJBV antifields: e.g., a covariant vector $A_m$ has a contravariant antifield $A^{*n}$, so the corresponding inner product involves only Kronecker deltas $\delta_m^n$.

Here we give a different interpretation of Strominger's result: Writing the open-string field as

$$\tilde{\Psi} = \left\{ \int_0^{\pi/2} d\sigma\, \tilde{\mathcal{Q}} + ... \right\} I$$

$$\tilde{\mathcal{Q}} = C^1(X'^m P_m + C'^0 B_0 + C'^1 B_1) + C^0(\tfrac{1}{2}g_{mn}X'^m X'^n + \tfrac{1}{2}g^{mn}P_m P_n + C'^0 B_1 + C'^1 B_0)$$

$$\langle g_{mn}(X) \rangle = \eta_{mn}$$

we interpret the field $g_{mn}$ as an *open* string field. In fact, if we compare with the usual oscillator expansion of $\Psi$, it is clear that this field (minus its vacuum value) is a combination of the usual open-string fields. (I.e., the linearized metric is an infinite sum, with appropriate coefficients, of the massive spin-2 fields that are mass eigenstates in the free theory.) Equivalently, it corresponds to eliminating the vacuum expectation values of all spin-2 tensors, except one, by simple field redefinitions: For any $\tilde{g}_{mn}$ (including traces of higher-spin fields) independent of $g_{mn}$ that has $\langle \tilde{g}_{mn} \rangle = k\eta_{mn}$, replace $\tilde{g}$ with a new field $\tilde{g}'_{mn} \equiv \tilde{g}_{mn} - kg_{mn}$, so $\langle \tilde{g}'_{mn} \rangle = 0$. Thus, if we start with many "metrics" (i.e., spin-2 fields with nonvanishing vacuum values), we can always redefine them so only one remains. (Similar remarks apply to gauge parameters, such as those for general coordinate transformations, where "vacuum value" refers then to the invariances of the vacuum, i.e., the global part of the local transformation.)

The consistency with Strominger's interpretation in terms of closed string fields follows from the fact that in the field theory of both open and closed strings the closed-open two-point vertex that follows from quantum corrections [6] implies a direct nonderivative two-point coupling between these open-string fields and the corresponding closed-string fields, like the nondiagonal mass-term type of coupling that can occur between two spin-2 fields [7], as here for the open-string field $g_{mn}$ and its closed-string analog. We will discuss this further in later sections. However, here we are considering a field theory of open strings only. This is actually more consistent with respect to an expansion in $\hbar$, since the relation of the closed-string self-coupling to the open-string one is $\hbar$-dependent. Closed-string states still appear in the theory, but only as bound states, so they are not described by fundamental fields, but rather by composite fields (just as, e.g., the hydrogen atom in QED). In this interpretation, the open-string field $g_{mn}$ introduced above is the only metric tensor available in the theory (the only spin-2 field which couples universally to the energy-momentum tensor of the open string). When one-loop corrections to its propagator are calculated, one finds a new massless spin-2 pole that didn't appear in the classical (tree-level) theory, i.e., the graviton.

At the linearized level it is already clear that the open-string field theory has gauge invariances for massless spin-2 fields that resemble those of massless spin-2 [8]. (In string field theory expansion in the ghost coordinates gives Stueckelberg fields.) The interacting case requires only a generalization of the Higgs mechanism to gravity. This can be derived in the same way as for massive spin-1, by performing a gauge transformation on a gauge non-invariant action, consisting of the gauge-invariant, massless, interacting action plus a mass term. The gauge parameter then becomes a new field. In the spin-2 case, the appropriate gauge parameter is the transformed coordinate. (This construction for massive spin-2 in string theory was mentioned in [9].) The gauge-invariant action is the Einstein-Hilbert action, and the mass term is the Fierz-Pauli term [10]

$$L_{FP} = -\tfrac{1}{4}M^2(h_{ab}^2 - h^a{}_a{}^2)$$

(which is the same as the mass term obtained from linearized gravity by dimensional reduction). The generalization of the mass term to the interacting case is somewhat ambiguous: In an expansion of the fields about their vacuum values, the constant term can always be canceled, the linear term must vanish (no cosmological term), and the quadratic term must fit the Fierz-Pauli term, but higher-order terms are arbitrary. In the model we write to study this sector of the string, we choose these terms to have as few derivatives as possible (a type of low-energy limit). Effectively, this means just defining

$$h^{ab} \equiv g^{mn}(\partial_m A^a)(\partial_n A^b) - \eta^{ab}$$

in $L_{FP}$ and throwing a $\sqrt{-g}$ in the measure.

The unique result (in D dimensions) is then:

$$S = \int d^D x \ \sqrt{-g}(R + M^2 L_M)$$

$$L_M = \tfrac{D(D-1)}{4} - \tfrac{D-1}{2} g^{mn}(\partial_m A^a)(\partial_n A^b)\eta_{ab}$$
$$+ \tfrac{1}{8} g^{mp} g^{nq} [(\partial_{[m} A^a)(\partial_{n]} A^b)][(\partial_{[p} A^c)(\partial_{q]} A^d)]\eta_{ac}\eta_{bd}$$

$$\langle g_{mn} \rangle = \eta_{mn}, \quad \langle A^a \rangle = x^a$$

This is exactly the Skyrme model [11] for the translation group. It has not only general coordinate invariance, under which $A^a$ transforms as scalars, but a separate, "internal", global Poincaré symmetry on $A^a$ [9]: $A^a \to A^b \lambda_b{}^a + \lambda^a$. (This is similar to a vierbein formalism with curved and flat indices, but here the tangent-space invariance is only global.) Unlike the analog for massive Yang-Mills, the quartic term is necessary here, and is required to be the Skyrme term to reproduce the Fierz-Pauli term. Unfortunately, this action has explicit flat-space metric factors $\eta_{ab}$, and it does not seem possible to eliminate them. (They can't be replaced with other tensor fields since after the above field redefinitions $g_{mn}$ is the only tensor with nonvanishing vacuum value. Similar remarks apply to eliminating the vacuum values of the other vectors $\tilde{A} \to \tilde{A}'$. In fact, all flat indices can be removed from everything except $A$ by using $\partial A$ as a vierbein: e.g., $\tilde{A}'^a \to \tilde{A}''^m \partial_m A^a$.) The string action consists of a sum of such terms for different masses $M$ (from the field redefinitions $\tilde{g} \to \tilde{g}'$ and $\tilde{A} \to \tilde{A}'$ at different mass levels): The sum for the mass term is less convergent than that for the $R$ term, and the sum of the coefficients might regularize to zero.

This suggests that the interpretation of the $\tilde{\Psi}^3$ action as background independent is singular. However, we can still interpret $g_{mn}$ (or perhaps all the spin-2 fields, to avoid singular field redefinitions) as a metric tensor, since it transforms covariantly under general coordinate transformations and has a nontrivial vacuum expectation value. Furthermore, because of the Stueckelberg fields supplied by string field theory, the action is general coordinate invariant. The interpretation is then that: (1) A metric tensor(s) and general coordinate invariance exist in the classical open-string field theory, but do not describe the graviton because of the Stueckelberg fields. (2) However, quantum contributions to the effective action cause this metric to describe a massless graviton (in addition to massive spin-2), with the dependence on the above tangent-space flat metric disappearing at low energy, near the graviton pole. Thus the metric tensor and general coordinate invariance are classical features of the theory, and it is only the massless pole that is the quantum feature.

Another interesting feature of the vacuum value of the open-string field is that it can be written as a pure abelian gauge transformation: In terms of the ghost-number operator $J$, the BRST operator $Q$ and left half of the BRST operator $Q_L$ can be expressed as:

$$\mathcal{J} \equiv C^0 B_0 + C^1 B_1 \quad \Rightarrow \quad [\mathcal{J}, Q] = \mathcal{Q}$$

$$\Rightarrow \quad [J, Q] = Q, \quad [J_L, Q] = Q_L; \quad J \equiv \int_0^\pi \mathcal{J}, \quad J_L \equiv \int_0^{\pi/2} \mathcal{J}$$

$$\Rightarrow \quad \langle \tilde{\Psi} \rangle = Q(-J_L I)$$

This is also true in ordinary gravity: In terms of the abelianized gauge transformation $\delta g_{mn} \approx \partial_{(m} \lambda_{n)}$, we have $\langle g_{mn} \rangle = \eta_{mn} = \partial_{(m} \frac{1}{2} x_{n)}$.

The origin of the bound-state massless graviton from massive spin-2 also explains the origin of the bound-state dilaton: The Stueckelberg formalism for massive spin-2 can be generalized from the above to include a scalar, representing massive spin-2 in terms of massless spins $2 \oplus 1 \oplus 0$, and all three of these fields appear automatically in open-string field theory [8]. In the massless limit of massive spin-2, the vector decouples but the scalar remains [12]: It is the dilaton.

## 3. THE GRAVITON IN FREE OPEN-STRING THEORY

The most unusual feature of closed-string bound-state generation in open-string theory is that it occurs at one loop, in constrast to more familiar bound-state mechanisms, such as the hydrogen atom, where the bound states, when represented by perturbation theory, are generated essentially by an infinite sum of ladder graphs, or the preon (random lattice) formulation of the string (or the QCD formulation of the hadronic string), where the string is generated as an infinite sum of all the leading graphs in the 1/N expansion. The reason is simple: The closed string is actually a bound-state of the *free* theory. This follows from the fact that any (amputated) one-loop propagator correction in an interacting theory is equivalent to the propagator of a composite field in the free theory: Consider any 3-point coupling $\chi^3$ (for simplicity we hide all derivatives, indices, etc.). The corresponding 1-loop $\chi$-propagator correction is then (in coordinate space) $\langle 0|J(x)J(y)|0\rangle$, where $J \sim \delta S/\delta\chi \sim \chi^2$ is the "current" coupling to $\chi$. In evaluating this 1-loop correction, we use the free propagator $\Delta_0(x-y) = \langle 0|\chi(x)\chi(y)|0\rangle$: $\langle 0|J(x)J(y)|0\rangle \sim [\Delta_0(x-y)]^2$. This is clearly the same as evaluating the propagator for the composite field $J$ in the free theory of the field $\chi$.

This bizarre mechanism is actually well known in two-dimensional field theory. In the Schwinger model (2D massless QED) [13], the photon propagator gets a scalar pole from the electron loop; it eats the scalar to become massive. The appearance of this scalar is now well understood: It is bosonization [14]. The vector current of the electron is equivalent to the gradient of a scalar field, even in the free theory of the electron. The photon couples to this current. This one-loop correction to the classical QED lagrangian ($F \equiv \epsilon^{ab}\partial_a A_b$)

$$L_0 = \tfrac{1}{2}F^2 + \bar\psi i\!\!\not{\partial}\psi + \bar\psi \!\!\not{A}\psi$$

can be expressed by introducing the bound-state field through the additional terms ($L = L_0 + \hbar L_1$)

$$L_1 = \tfrac{1}{2}\phi\Box\phi + \phi F$$

resulting from the bosonization of $\psi$. We have introduced $\phi$ as a redundant field to $\psi$: Introducing a separate field for a bound state is always redundant (but not incorrect, although coupling constants need to be modified because of double counting). Integrating out $\phi$ classically (tree graphs) produces a term $-\tfrac{1}{2}F\Box^{-1}F$ in the effective action, the same as integrating out $\psi$ at one loop.

We therefore have three two-dimensional field theories that describe this bound-state phenomenon: (1) In the theory of a free, massless fermion, $L = \bar\psi i\!\!\not{\partial}\psi$, a composite field describing the scalar bound state is *defined* by the usual bosonization formula $\bar\psi\gamma_a\psi \equiv \epsilon_a{}^b\partial_b\phi$. The consistency of such a definition follows from just current conservation (in any dimension), but the fact that the propagator of the composite field $\phi$ has poles follows from kinematics: Classical massless particles in D=2 travel at the speed of light to either the left or right. $\gamma_\pm$ picks out the components of $\psi$ and $\bar\psi$ for electron and positron both traveling to the left or both to the right (as follows from their equations of motion $\gamma_\pm\partial_\mp\psi = 0$). Two massless particles starting at the same point and traveling in the same direction are never separated, and therefore act as a bound state. (2) In massless QED ($L = L_0$), the photon couples to this composite field, so at one loop the bound state shows up in the propagator of the fundamental field $A_a$. (3) In the effective field theory described by $L = L_0 + \hbar L_1$, $\phi$ is a fundamental field rather than a composite one, but the physics is identical to that described by just $L_0$. If $\phi$ is eliminated by its equation of motion, we get back the formulation in terms of just $L_0$, but including the one-loop contribution that shows the presence of the bound state. Although the free fermion theory is sufficient to describe the bound state, the interacting theory of QED automatically points out the existence of the bound state by showing it in the one-loop propagator of a fundamental field, and the effective field theory helps to elucidate the mechanism by which this happens.

A similar mechanism occurs for the string: An open-string current is equivalent to a closed string, even in the free theory of the open string. This current can be found from the $\Psi^3$ interaction of the interacting theory. As for the Schwinger model, where the bound state can be described as a composite (quadratic) field in the theory of a free fermion (bosonization), the closed-string bound state can be described as a composite (quadratic) field in the theory of a free open string, and our discussion of the interacting open-string field theory and (in the following section) the interacting field theory of open and closed strings is for pedagogical purposes (and because those theories are interesting in their own right).

The Schwinger-model/string analogy is then

| Schwinger model | string field theory |
|---|---|
| photon $A_a$ | open-string metric $g_{mn}$ |
| electron $\psi$ | rest of open-string field $\Psi$ |
| scalar bound state $\phi$ | closed-string bound state $\Phi$ |
| current $\bar\psi \gamma_a \psi$ | current $\Psi * \Psi$ |
| bosonization $\bar\psi \gamma_a \psi = \epsilon_a{}^b \partial_b \phi$ | $\Psi * \Psi = \Upsilon^\dagger G \Phi + ...$ |

where $\Upsilon$ is the open-closed 2-point vertex operator [6] (in our notation, it takes an open-string field to a closed-string field), and $G$ takes care of the zero-modes associated with global $\sigma$-translation invariance on the world sheet:

$$G \equiv b_1 \delta(\Delta N), \quad b_1 \equiv \int d\sigma\, B_1, \quad \Delta N \equiv \{Q_\Phi, b_1\}$$

Unlike the Schwinger model, $\Upsilon$ contains no spacetime derivatives, so we can locally invert the expression for the current in terms of the closed-string field as

$$\tilde\Phi \equiv b_0 \Upsilon b_0 \Psi * \Psi \approx \Pi \Phi, \quad \Pi \equiv b_0 \Upsilon b_0 \Upsilon^\dagger G$$

Π is essentially the product of two $\delta$-functionals that projects out open-string states that can couple to closed strings. The ghost insertions $b_0$ are gauge dependent, as expected from the fact that $\Phi$ has a gauge parameter $\Lambda_\Phi$ independent of $\Psi$'s gauge parameter $\Lambda_\Psi$. They are exactly those that appear in the iteration of the lowest-order propagator corrections expressed by the open-closed coupling $\Phi^\dagger G \Upsilon \Psi$:

$$\Delta_\Psi = \frac{b_0}{p^2 + M_\Psi^2}, \quad \Delta_\Phi = \frac{b_0 c^1}{p^2 + M_\Phi^2}$$

$$\Rightarrow \quad \Delta_\Phi \cdot G\Upsilon \cdot \Delta_\Psi \cdot \Upsilon^\dagger G \cdot \Delta_\Phi ... \Upsilon^\dagger G \sim c^1 G \Pi \Pi ... \Pi$$

Thus $\tilde{\Phi}$ is just a field redefinition of $\Phi$ (at least for the propagating components).

The analogy also relates to the interpretation of the previous section: For the Schwinger model in terms of fields $A$, $\psi$, and $\phi$ ($L_0 + \hbar L_1$), it is $\phi$ that has the physical bosonic polarization, which shows up in the $A$ propagator only because of its coupling to $\phi$. On the other hand, for the formulation in terms of just the fields $A$ and $\psi$ ($L_0$), the electromagnetic field, which classically has no physical polarizations (in D=2), has gained a physical polarization at one loop. For the string in terms of $\Psi$ and $\Phi$, it is $\Phi$ that describes the graviton; but in terms of just the open-string field $\Psi$, the metric tensor(s) $g_{mn}$ in $\Psi$ describes only massive spin-2 classically but develops a (massless) graviton pole (as well as a dilaton pole) at one loop.

## 4. OPEN-CLOSED STRING FIELD THEORY

Unlike the Schwinger model, where the field $\psi$ that generates the bound state has no self-interactions, the self-interactions of the open string $\Psi$ generate closed-string (bound-state) self interactions at higher loops, so our analysis in terms of an effective action of both open and closed strings requires an analysis beyond one loop. In string theory the order in $\hbar$ of a graph is related to the Euler number: In field theory language, the contributions to the effective action (appearing as $e^{S/\hbar}$) are of the form, for arbitrary nonnegative $m$ and $n$,

$$\Psi^{m+1}(\hbar\Phi)^n, \quad \Phi(\hbar\Phi)^{n+1}$$

plus terms of this form but higher order in $\hbar$. As in ordinary field theory, $\hbar$ and the coupling constant $g$ appear in $S/\hbar$ only in the combination $\hbar g^2$. There might also be a tadpole term $\hbar\Phi$, but not in the supersymmetric theory. Again, the normal $\hbar$ counting, with $\hbar$'s appearing only in loops, is consistent only in the formulation where only $\Psi$ appears as a fundamental field. The field theory of closed strings was developed in [15]. Field theory of open and closed strings has been discussed in [16], but in a form where the existence of gauge invariance was unclear.

For our purposes it will be sufficient to consider adding to the classical open-string lagrangian
$$L_0 = \tfrac{1}{2}\Psi^\dagger Q_\Psi \Psi + \tfrac{1}{3}\Psi^\dagger \Psi * \Psi$$
the effective closed-string terms ($L \approx L_0 + \hbar L_1 + \hbar^2 L_2$)
$$L_1 = \tfrac{1}{2}\Phi^\dagger G Q_\Phi \Phi + \Phi^\dagger G \Upsilon \Psi + \Phi \Psi^{n+2}\text{-terms} + \tfrac{1}{2}\Psi^\dagger V_\Psi \Psi$$
$$L_2 = \Phi^3\text{-term} + \Phi^2 \Psi^{n+1}\text{-terms} + \tfrac{1}{2}\Phi^\dagger G V_\Phi G \Phi$$
We could replace $\Phi$ with the field $\hat{\Phi} \equiv G\Phi$ satisfying the constraints $b_1 \hat{\Phi} = \Delta N \hat{\Phi} = 0$, which would simply absorb all $G$'s, except for the replacement $\tfrac{1}{2}\Phi^\dagger G Q_\Phi \Phi \to -\tfrac{1}{2}\hat{\Phi}^\dagger c^1 Q_\Phi \hat{\Phi}$. This field is more useful for conformal field theory because its Hilbert space more closely corresponds to the direct product of two open-string Hilbert spaces. Its use is analogous to the use of chiral superfields in supersymmetry: The term $-\tfrac{1}{2}\hat{\Phi}^\dagger c^1 Q_\Phi \hat{\Phi}$ could then be expressed without the $c^1$ as integrated over the "chiral" subspace, whereas all other terms, including all quantum corrections, would be integrated over the full space. However, we prefer to work with unconstrained fields.

Like $*$ and $I$, $\Upsilon$ is a $\delta$-functional, times ghost factors that follow from including the ghost coupling to the world-sheet curvature: If we write the $\Phi^\dagger G \Upsilon \Psi$ term as a double integral, with $\Phi^\dagger G$ evaluated on closed-string coordinates and $\Psi$ evaluated on open-string coordinates (again using $C^0$ and $C^1$ as the ghost coordinates), $\Upsilon$ is a $\delta$-functional of the closed minus the open coordinates, times a factor $C^0(0)$, where $\sigma = 0$ is the interaction point (where the world-sheet curvature is). This follows from considering ghost zero-modes: For physical fields, $\Psi$ and $G\Phi$ are independent of them; the open-string coordinate integration integrates over one, while the closed integrates over two. This leaves three for $\Upsilon$: two for the $\delta$-functional, one for $C^0(0)$. The counting is different for bosonized ghost coordinates. In this form $\Upsilon$ introduces no spacetime derivatives, in contrast to the Schwinger model. The exact vertex $\hat{\Upsilon}$ that appears in the one-loop diagram [6] as the effective action contribution $\Psi^\dagger \hat{\Upsilon}^\dagger G \Delta_\Phi G \hat{\Upsilon} \Psi$ differs from $\Upsilon$ by a conformal transformation that includes derivatives, which is just an invertible field redefinition.

$\Upsilon$ essentially projects out the open-string states that couple directly to the closed-string states (i.e., the group theory singlets, like the open-string metric $g_{mn}$), in analogy to the way that $\gamma_a$ in the Schwinger model projects out the chiral part of $\psi \otimes \bar\psi$. Furthermore, whereas in the Schwinger model integrating out the bound-state field $\phi$ gives an identical result to a fermion loop with two external lines, integrating out the closed-string field gives at best only the part of the nonplanar one-loop graph containing the physical closed-string poles. This is probably related to the fact that the bosonization expression for $\phi$ in terms of $\psi$ is invertible, while the expression of $\Phi$ in terms of $\Psi$ might not be invertible to express the open-string field in terms of the closed string. Therefore, we include terms $\frac{1}{2}\Psi^\dagger V_\Psi \Psi$ and $\frac{1}{2}\Phi^\dagger G V_\Phi G \Phi$ representing the local parts of one-loop propagator corrections (corresponding to contracting the lengths of the propagators in terms of proper/world-sheet time), while the $\Phi^\dagger G \Upsilon \Psi$ term incorporates the (physical) pole parts of those corrections.

For studying the effective action, we can concentrate on the quadratic terms (dropping the $\hbar$'s):

$$L \approx \tfrac{1}{2}\Xi^\dagger \tilde G Q \Xi, \quad \tilde G Q = Q^\dagger \tilde G, \quad \delta\Xi \approx Q\Lambda, \quad Q^2 = 0, \quad \Xi = \begin{pmatrix} \Psi \\ \Phi \end{pmatrix}, \quad \tilde G = \begin{pmatrix} 1 & 0 \\ 0 & G \end{pmatrix}$$

$$Q = Q_0 + Q_1 + Q_2 = \begin{pmatrix} Q_\Psi & 0 \\ 0 & Q_\Phi \end{pmatrix} + \begin{pmatrix} 0 & \Upsilon^\dagger G \\ \Upsilon & 0 \end{pmatrix} + \begin{pmatrix} V_\Psi & 0 \\ 0 & V_\Phi G \end{pmatrix}$$

$\Psi, \Phi, Q_0, G, \Upsilon, V_\Psi, V_\Phi$ have ghost numbers $-\frac{1}{2}, 0, 1, -1, \frac{3}{2}, 1, 2$, respectively, in terms of the ghost number operator that is antihermitian with respect to the Hilbert-space metric; i.e., the integration measure is defined to have vanishing ghost number.

Unlike the $A\phi$ coupling of the Schwinger model, the open-closed coupling is not gauge invariant under the independent gauge transformations $\Lambda = (\Lambda_\Psi, \Lambda_\Phi)$ of the uncoupled open and closed string field theories: The string action $S = \int L$ is actually the ZJBV operator, which includes not only the classical gauge-invariant action but also the BRST (gauge) transformations. (In the quadratic part of the action, this is simply the statement that $Q$ is both the kinetic operator and the generator of gauge transformations.) Through their dependence on $c^0$, the string fields contain both the usual fields and their ZJBV antifields: In the expansion $\Psi = \psi_+ + c^0 \psi_-$, $\psi_+$ contains only fields while $\psi_-$ contains only antifields. The usual ZJBV antibracket of fields and antifields then follows from the string field antibracket $(\Psi^\dagger, \Psi) = \delta$ [17], in terms of a $\delta$-functional of all the coordinates, including $c^0$. The closed-string field behaves similarly, except $(\Phi^\dagger, \Phi) = c^1 \delta$ has the factor $c^1$ because the $c^1$-independent part of $\Phi$ drops out of the action. (Equivalently, $(\hat\Phi^\dagger, \hat\Phi) = G\delta$, where $G$ is analogous to the factor appearing in the definition of functional differentiation of chiral superfields to maintain the constraints on the field.) Gauge invariance is then the usual statement $(S, S) = 0$. The term $\Phi^\dagger \Upsilon G \Psi$ ($Q_1$) thus contributes a crossterm not only to the action, but also to the gauge transformations. (We consider an effective action calculated by background field gauge methods, so gauge fixing does not break the gauge invariance of the effective action.)

We now examine the gauge invariance condition $Q^2 = 0$ perturbatively. (This can be considered an $\hbar$ expansion if we redefine $\Psi \to \hbar^{1/4}\Psi$, $\Phi \to \hbar^{-1/4}\Phi$, followed by $\hbar \to \hbar^2$: Then $Q = Q_0 + \hbar Q_1 + \hbar^2 Q_2$.) Of the resulting relations

$$Q_0^2 = \{Q_0, Q_1\} = Q_1^2 + \{Q_0, Q_2\} = \{Q_1, Q_2\} = Q_2^2 = 0$$

the first two expressions are known to vanish. The middle identity states that $Q_1^2$ is BRST trivial. In the open-string sector, this follows from the fact that it is BRST invariant but not in the (operator) BRST cohomology: Otherwise $Q_1^2|0\rangle$ would be a Poincaré invariant state in the cohomology, but there are none of that ghost number. Since on shell and translation invariant means massless, the only on-shell zero-momentum scalars in the cohomology are those of Yang-Mills theory: the Yang-Mills ghost at zero momentum $|0\rangle$ (corresponding to the identity operator) and its antifield $CC'C''|0\rangle$ (corresponding to the operator $CC'C''$ with ghost number 3), whereas $\Upsilon^\dagger G\Upsilon$ has ghost number 2. Thus, $\Upsilon^\dagger G\Upsilon = -\{Q_\Psi, V_\Psi\}$ for some $V_\Psi$. In the closed-string sector $\Upsilon\Upsilon^\dagger$, in geometrical terms, describes a closed string that breaks into an open string and instantaneously reconnects into a closed string. In other words, the cylindrical world-sheet describing the spacetime propagation of the closed string has an infinitesimal hole. Because of the ghost insertions $C^0(0)$ in $\Upsilon$ and $\Upsilon^\dagger$ being multiplied at the point of that hole resulting in the squaring of an anticommuting c-number, this diagram vanishes. We can thus set $V_\Phi = 0$. (The same result holds if we use $\hat{\Upsilon} = U_\Phi \Upsilon U_\Psi$ instead of $\Upsilon$, since it differs only by a unitary transformation $U$: $\hat{\Upsilon}\hat{\Upsilon}^\dagger = U_\Phi \Upsilon \Upsilon^\dagger U_\Phi^{-1} = 0$.) The remaining identities then become $\Upsilon V_\Psi = V_\Psi^2 = 0$, which we expect to follow also from squaring of fermionic insertions. There are additional $V_\Psi$ and $V_\Phi$ terms in nonorientable string theories (such as the SO(32) superstring) [18].

At the free level, this calculation should be identical (in an appropriate representation of the ghost coordinates) to the calculation of closure of the Lorentz algebra in the light-cone theory: At the quadratic level the BRST operator derived by covariantizing [19,20] the nontrivial light-cone Lorentz generators [21] is generally identical to that obtained by the usual BRST methods (in this case, with $\Upsilon$ and not $\hat{\Upsilon}$ as the open-closed vertex). It has been shown [22] that the Lorentz algebra closes for open-closed bosonic light-cone string field theory [23], although there the cancellation of the $\Upsilon^\dagger G\Upsilon$ term was with an "anomalous" term coming from a nonplanar loop diagram generated by the squaring of the $\Psi^3$ term. However, the generation of the closed-string bound state from the nonplanar graph in the light-cone string field theory (if it occurs at all) is different than in both the usual covariant one and the covariantized light-cone one. Still, it should be possible to obtain the analog of our result for the light cone by adding a corresponding local term to the "classical" lagrangian and subtracting the same term from the one-loop corrections. In any case, we will assume the $V_\Psi$ term is sufficient to solve the problem (as implied by our open-string cohomology argument), which does not require consideration of loop diagrams to show the gauge invariance of the free theory. In the following section this assumption will be supported by examples from finite subsets of the open- and closed-string fields. There we will give a general analysis of quadratic lagrangians exhibiting nonderivative couplings between two different sectors, of which open-closed string theory should be a special case.

## 5. GENERAL ANALYSIS OF FIELD MIXING

Rather than delving into the technical details of this quadratic coupling between fields of the open and closed strings, we analyze this phenomenon in a more general setting. We first consider the most general possible (relativistic) free lagrangian (up to field redefinitions) for any set of fields. This result is known from first-quantized BRST methods [20]:

$$L = \tfrac{1}{2}\xi^\dagger Q\xi; \quad Q = \hat{Q}_0 + iMM^{cm} + M^2 c, \quad \hat{Q}_0 = cp^2 + iM^{ca}p_a - \tfrac{1}{2}bM^{cc}$$

where $M^{ij}$ are the D-dimensional generators of OSp(D,1|2), an extension of the Lorentz group to include ghosts. The index $c$ labels one of the two ghost directions and $m$ labels the single reduced dimension used in the generation of mass by dimensional reduction: $p_m = M$. ($c$ and $b$ are $c^0$ and $b_0$ for the string.) We have explicitly separated out the mass dependence, writing the general BRST operator $Q$ in terms of the massless BRST operator $\hat{Q}_0$. Irreducible representations of the OSp(D−1,1|2) subgroup generated by $M^{ij}$ for $i,j \neq m$ are representations of the massless version of this extended Lorentz group, consisting of a single gauge field and its ghosts (and antifields, through dependence on $c$): Such representations can eat each other to become massive representations of OSp(D,1|2).

We next consider the most general way to break up $Q = Q_0 + \hbar Q_1 + \hbar^2 Q_2$ such that $Q^2 = 0$ perturbatively in $\hbar$. Since $Q$ has the most general form, $Q_0$ can differ only in how the operators $M$ and $M^{ij}$ are represented. However, $M^{ij}$ for $i, j \neq m$ (the OSp(D−1,1|2) subgroup) must be represented in the same way to preserve the Lorentz subgroup SO(D−1,1); thus their $\hat{Q}_0$ terms are the same. In the light-cone string theory, from which the covariant result can be derived, this is the statement that the generators of SO(D−2) are the naive free first-quantized operators, and get no quantum corrections, while the remaining Lorentz generators include terms from the open-closed coupling, as well as other interaction terms. Since each representation of OSp(D,1|2) (and in particular $M^{cm}$) requires an appropriate set of fields, its representation in $Q_0$ can differ from that in $Q$ only by mixing between OSp(D−1,1|2) representations of the same type. However, it seems that the only way to break up an individual OSp(D,1|2) representation that is consistent with satisfying $Q^2 = 0$ perturbatively is the separation of its fields into those with odd and even numbers of "m" indices. We therefore restrict the massless representations of $Q$ (i.e., the subset or our original fields with zero eigenvalue of the operator $M$) to be only of the form of an "even" or "odd" part of a representation of OSp(D,1|2), so they can mix with the corresponding parts of the massive representations. This holds for the massless sectors of the string: the vector and antisymmetric tensor are already the even parts of the corresponding massive representations; the graviton and the dilaton together correspond to the even part of massive spin 2. Furthermore, for the string the separation into odd and even numbers of $m$ indices corresponds to separation into odd and even numbers of oscillators: In $Q_0$, $MM^{cm}$ is cubic in oscillators while $M^2$ is quadratic; in $Q_1$, the $MM^{cm}$ ($c$-independent) term in the operator $\Upsilon$ is odd in oscillators while the $M^2$ term is even (as follows from the representation of the overlap integrals used in evaluating $\delta$-functionals as Gaussians).

The most general form of a representation $\mathcal{Q}$ of the $iMM^{cm}$ term in any of the pieces $Q_0$, $Q_1$, $Q_2$ is then

$$\mathcal{Q} = \alpha \mathcal{A} + \alpha^\dagger \mathcal{A}^\dagger, \quad iM^{cm} = \mathcal{A} + \mathcal{A}^\dagger$$

where we have broken up the original $M^{cm}$ in $Q$ into the piece $\mathcal{A}$ that takes the even part of an OSp(D,1|2) representation into the odd part, and the piece $\mathcal{A}^\dagger$ that takes odd into even. $\alpha$ is a matrix that mixes the different copies of an even part and maps them to the different copies of the corresponding odd part (and similarly for $\alpha^\dagger$ mapping odd to even). Thus, $Q_0$, $Q_1$, and $Q_2$ are block diagonal with respect to OSp(D,1|2): The fields for any one block consist of all the copies of any particular OSp(D,1|2) representation (but of different masses), plus all the massless fields corresponding to the even or odd part of that representation. Restricting our attention to such a set of fields for a particular representation of OSp(D,1|2), we can divide the

fields into even and odd as $\xi = (\xi_{e\epsilon}, \xi_{d\delta})$, where "$e$" labels the copies of the even part and "$\epsilon$" labels the components of an even part, and similarly for the odd stuff $d\delta$ (but $d$ has a different range than $e$, and $\delta$ different than $\epsilon$). Then the index notation for the above equation is

$$\begin{pmatrix} \mathcal{Q}_{e\epsilon e'\epsilon'} & \mathcal{Q}_{e\epsilon d'\delta'} \\ \mathcal{Q}_{d\delta e'\epsilon'} & \mathcal{Q}_{d\delta d'\delta'} \end{pmatrix} = \begin{pmatrix} 0 & \alpha^\dagger{}_{ed'}\mathcal{A}^\dagger{}_{\epsilon\delta'} \\ \alpha_{de'}\mathcal{A}_{\delta\epsilon'} & 0 \end{pmatrix}$$

Solving $Q^2 = 0$ perturbatively then completely determines the $Q$'s in terms of the $\alpha$'s (and requires that $Q_3$ have no $\mathcal{Q}$ term):

$$Q_0 = \hat{Q}_0 + \mathcal{Q}_0 + \mu_0 c, \quad Q_1 = \mathcal{Q}_1 + \mu_1 c, \quad Q_2 = \mu_2 c$$

$$\mathcal{Q}_k = \alpha_k \mathcal{A} + \alpha^\dagger{}_k \mathcal{A}^\dagger$$

$$\mu_0 = \alpha_0 \alpha^\dagger{}_0 + \alpha^\dagger{}_0 \alpha_0, \quad \mu_1 = \alpha_{(0}\alpha^\dagger{}_{1)} + \alpha^\dagger{}_{(0}\alpha_{1)}, \quad \mu_2 = \alpha_1 \alpha^\dagger{}_1 + \alpha^\dagger{}_1 \alpha_1$$

Lastly, we break up our fields $\xi$ into two sets $\psi$ and $\phi$, which will be the analog of the open- and closed-string fields. This division of fields is chosen to preserve even and odd pieces: For any representation of OSp(D,1|2), any linear combination of the copies of the even part (including those of different mass) may go into $\psi$, and any linear combination of the odd; the independent combinations go into $\phi$. In particular, we may break up an OSp(D,1|2) representation so some of its OSp(D−1,1|2) representations are in $\psi$ and some are in $\phi$: i.e., the gauge field and its even Stueckelberg fields are in one place, the rest of its Stueckelberg fields in the other. Finally, we choose an $\alpha_0$ that is diagonal with respect to this $\psi/\phi$ breakup, and an $\alpha_1$ that is off-diagonal. That makes $Q_0$ and $Q_2$ diagonal, $Q_1$ off-diagonal. We expect this description to be equivalent to that just given for the string, as we now demonstrate with two examples.

For our first example we consider the generation of mass by dimensional reduction, and choose the above perturbation expansion as simply an expansion in the mass: i.e., we define our separation of fields into those with odd and even numbers of $m$ indices, so that $\alpha_0 = 0$ and $\alpha_1 = M$:

$$Q_0 = \hat{Q}_0, \quad Q_1 = iMM^{cm}, \quad Q_2 = M^2 c$$

The lagrangian is then

$$L = \tfrac{1}{2}\psi^\dagger(\hat{Q}_0 + M^2 c)\psi + i\phi^\dagger MM^{cm}\psi + \tfrac{1}{2}\phi^\dagger(\hat{Q}_0 + M^2 c)\phi$$

For the special case of the antisymmeric tensor $A_{ab}$ (where dimensional reduction gives also the vector $A_{ma}$), this agrees with results for the massless level of the U(N) bosonic string [6], where the antisymmetric tensor from the closed string mixes with the (singlet part of the) vector from the open string [24] through a term $B^a A_{ma}$ involving $A_{ab}$'s Nakanishi-Lautrup (NL) field $B_a$.

We next eliminate $\phi$ by its equation of motion, inverting $\hat{Q}_0 + M^2 c$ in the usual gauge to get the propagator $b/(p^2 + M^2)$ (and using the identity $(M^{cm})^2 = -\frac{1}{2}M^{cc}$) to find the equivalent lagrangian in terms of just $\psi$

$$\tilde{L} = \tfrac{1}{2}\psi^\dagger \tilde{Q}\psi$$

$$\tilde{Q} = \hat{Q}_0 + M^2 c + \tfrac{1}{2}M^2 \frac{b}{p^2+M^2}M^{cc} = \hat{Q}_0 + M^2(c + \tfrac{1}{2}\frac{b}{p^2}M^{cc}) + \mathcal{O}(M^4) \equiv \tilde{Q}_0 + \tilde{Q}_1 + ...$$

$\tilde{Q}_0 + \tilde{Q}_1$ is the result that would be obtained for $\tilde{Q}$ if the term $\tfrac{1}{2}\phi^\dagger M^2 c\phi$ were ignored (as in string theory, where the analogous $V_\Phi$ term does not occur). Since $\{\tilde{Q}_0, \tilde{Q}_1\} = 0$, $\tilde{Q}_1$ represents a correction to the lagrangian consistent with global BRST invariance to that order in $M^2$. However, the complete $\tilde{Q}$ is necessary for total BRST invariance, and the $\tfrac{1}{2}\phi^\dagger M^2 c\phi$ term is necessary in $L$ for gauge invariance. (Although in string theory the analogous term is unnecessary, the corresponding term for $\psi$ is.) This model is unlike the full string in that $Q_1$ for the string ($\Upsilon$) includes $c$ ($c^0$) terms, which contribute to the propagator as defined by $L = L_0 + L_1 + L_2$ in the $b_0 = 0$ gauge. $c$ terms in $Q$ contribute terms to the gauge-invariant action containing just physical fields, while other terms in $Q$ contribute terms containing NL auxiliary fields. However, in the calculation of [6] for the massless sector of the U(N) bosonic string, the portion of $\Upsilon$ actually used was exactly the $M^{cm}$ used here, and the only part of $\tilde{Q}_1$ obtained by explicit calculation was exactly the $bM^{cc}/p^2$ obtained here, while the $c$ term (analogous to $V_\Psi$) was inferred to also follow from $\Upsilon^\dagger G\Delta_\Phi G\Upsilon$ by BRST invariance of $\tilde{Q}_1$.

We now discuss an example that exhibits direct coupling between physical fields of the same spin, but different masses (as opposed to the previous example, which illustrated coupling between particles of the same mass, but coupled a physical field to an NL field). In the more interesting cases one particle is massless, so we'll assume that restriction here for convenience. In the special case of spin 2, this is analogous to the coupling between the massive fields of the open string and the graviton (and dilaton) of the closed string, on which we focused in section 2. As for the case of spin 1 ("vector meson dominance"), such actions are obtained simply by (1) writing the kinetic terms for two particles of the same spin, but different masses, (2) coupling both (gauge covariantly) to two "matter" sectors, with different couplings for the different sectors, and (3) making field redefinitions that diagonalize the couplings to the two different matter sectors, which makes the gauge fields' mass terms off-diagonal. For example, for the spin-1 case, we start with the usual free lagrangians for a massless photon $A'$ and massive rho-meson $\rho'$, couple $A'+\rho'$ to strongly interacting matter and $A'-\rho'$ to weakly interacting matter, and then redefine $\rho = A'+\rho'$ and $A = A'-\rho'$ as our new fields, so the mass term $\rho'^2 \sim (\rho-A)^2$ has cross terms. (In the spin-2 case, the graviton's coupling is nonabelian, so we have to be more careful with interactions [7], but we'll again restrict ourselves here to just the quadratic part of the lagrangian.)

We begin by dividing up the fields of the massive states into $\psi'$, with even numbers of $m$ indices, and $\chi$, with odd. In the case of spin 1, $\psi'$ is a vector while $\chi$ is a scalar; for spin 2, $\psi'$ is a tensor plus a scalar, while $\chi$ is a vector. We next write the lagrangian

$$L = \tfrac{1}{2}\psi'^\dagger(\hat{Q}_0 + m^2 c)\psi' + \tfrac{1}{2}\chi^\dagger(\hat{Q}_0 + m^2 c)\chi + m\chi^\dagger \mathcal{A}\psi' + \tfrac{1}{2}\phi'^\dagger \hat{Q}_0 \phi' = \tfrac{1}{2}\xi'^\dagger Q' \xi'$$

$$\xi' = \begin{pmatrix} \psi' \\ \chi \\ \phi' \end{pmatrix}, \quad Q' = \begin{pmatrix} \hat{Q}_0 + m^2 c & m\mathcal{A}^\dagger & 0 \\ m\mathcal{A} & \hat{Q}_0 + m^2 c & 0 \\ 0 & 0 & \hat{Q}_0 \end{pmatrix}$$

where $m$ is the value of the mass for the massive states. To mix the massless $\phi'$ with $\psi'$ in a way that preserves the matrix structure, we must choose it to be the same representation of OSp(D$-$1,1|2). Thus, for spin 2 $\phi'$ must include the dilaton as well as the graviton. (Of course, it is still possible to produce such couplings between massive and massless spin 2 without a dilaton [7], but the dilaton is necessary for $Q^2 = 0$ to be satisifed order-by-order in the perturbation expansion $Q = Q_0+Q_1+Q_2$.)

We then "undiagonalize" by

$$\psi' = \frac{1}{\sqrt{2}}(\psi + \phi), \quad \phi' = \frac{1}{\sqrt{2}}(\psi - \phi) \quad \Rightarrow \quad L = \tfrac{1}{2}\xi^\dagger Q \xi, \quad Q = Q_0 + Q_1 + Q_2$$

$$Q_0 = \begin{pmatrix} \hat{Q}_0 + \tfrac{1}{2}m^2 c & \tfrac{1}{\sqrt{2}}m\mathcal{A}^\dagger & 0 \\ \tfrac{1}{\sqrt{2}}m\mathcal{A} & \hat{Q}_0 + \tfrac{1}{2}m^2 c & 0 \\ 0 & 0 & \hat{Q}_0 \end{pmatrix}$$

$$Q_1 = \begin{pmatrix} 0 & 0 & \tfrac{1}{2}m^2 c \\ 0 & 0 & \tfrac{1}{\sqrt{2}}m\mathcal{A} \\ \tfrac{1}{2}m^2 c & \tfrac{1}{\sqrt{2}}m\mathcal{A}^\dagger & 0 \end{pmatrix}, \quad Q_2 = \begin{pmatrix} 0 & 0 & 0 \\ 0 & \tfrac{1}{2}m^2 c & 0 \\ 0 & 0 & \tfrac{1}{2}m^2 c \end{pmatrix}$$

(We could have obtained a mass other than $\tfrac{1}{\sqrt{2}}m$ by using a different undiagonalization and a correspondingly modified splitting of $Q$ into $Q_0$, $Q_1$, and $Q_2$.) $Q_0$ gives the standard lagrangian for the decoupled fields (with masses $\tfrac{1}{\sqrt{2}}m$ and 0) in BRST form [20]. Now $Q_1$ contains not only the $MM^{cm}$ term of the previous example, but also a $c$ term that gives the desired direct coupling between physical fields. Examination of the explicit $\Upsilon$ of [6] shows that these are exactly the terms it contains that couple the first massive level of the open string to the fields of the closed string describing the graviton and dilaton: To the gauge-invariant action the $MM^{cm}$ term contributes $B^a A_a$ (in terms of the NL field $B_a$ of the graviton and linearized Stueckelberg field $A_a$ of massive spin 2) and $B\varphi$ (in terms of the NL field $B$ of $A_a$ and the dilaton $\varphi$), while the $c$ term contributes $h^{ab}\tilde{h}_{ab}$ (in terms of the linearized graviton $h_{ab}$ and linearized massive spin 2 $\tilde{h}_{ab}$) and $\varphi\tilde{\varphi}$ (in terms of the other Stueckelberg field $\tilde{\varphi}$ of massive spin 2). ($Q_2$ gives further diagonal mass terms $A_a^2$, $h_{ab}^2$, and $\varphi^2$.) We can also recognize this as a special case of the general mixing described above (which gives a more direct derivation of this result), given by $\alpha_0 = \frac{1}{\sqrt{2}}m(1,0)$, $\alpha_1 = \frac{1}{\sqrt{2}}m(0,1)$.

## ACKNOWLEDGMENT

I thank Martin Roček, Fred Goldhaber, and Peter van Nieuwenhuizen for discussions. This work was supported in part by the National Science Foundation Grant No. PHY 9309888.